\RequirePackage{ifpdf} 
\ifpdf
\documentclass[pdftex,12pt,a4paper]{article}  
\else
\documentclass[12pt,a4paper]{article} 
\fi

\parskip=\medskipamount

\usepackage[mathscr]{euscript} 	
\usepackage{bbm} 
\usepackage{amsfonts,amsmath,amssymb}
\ifpdf
\usepackage[pdftex]{graphicx}	
\usepackage[usenames,pdftex]{color}		
\usepackage[pdftex,bookmarks]{hyperref}
\hypersetup{bookmarksopen,bookmarksnumbered,
		pdfnewwindow=true,
		pdfauthor = {Simon J. Tyler},
		pdftitle = {Comments on the complex linear Goldstino superfield},
		pdfsubject = {},
		pdfkeywords = {Supersymmetry, Goldstino},
		pdfcreator = {LaTeX with hyperref package},
		pdfproducer = {pdflatex}
}
\else
\usepackage[dvips]{graphicx}
\usepackage[usenames,dvips]{color}
\usepackage[dvips,bookmarks,colorlinks=false]{hyperref}
\fi
\def\a{\alpha}
\def\b{\beta}
\def\c{\chi}

\def\f{\phi}
\def\g{\gamma}

\def\j{\psi}
\def\k{\kappa}
\def\l{\lambda}
\def\m{\mu}

\def\q{\theta}
\def\r{\rho}
\def\s{\sigma}

\def\F{\Phi}

\def\S{\Sigma} 

\newcommand{\da}{{\dot{\alpha}}}
\newcommand{\db}{{\dot{\beta}}}

\newcommand{\ada}{{\alpha\dot\alpha}}

\def\ds1{\ensuremath{\mathbbm{1}}}
\newcommand{\rmd}{{\rm d}}
\newcommand{\rme}{{\rm e}}
\newcommand{\rmi}{{\rm i}}

\newcommand{\cD}{{\cal D}}
\newcommand{\cDb}{\bar\cD}

\newcommand{\cN}{{\cal N}}
\newcommand{\st}{{\tilde\s}}

\newcommand{\pd}{\partial}

\def\const{{\rm const}}

\def\intx{\int\!\!{\rmd}^4x\,}

\newcommand{\be}{\begin{equation}}
\newcommand{\ee}{\end{equation}}
\newcommand{\bea}{\begin{eqnarray}}
\newcommand{\eea}{\end{eqnarray}}
\newcommand{\non}{\nonumber}
\newcommand{\bm}[1]{\mbox{\boldmath$#1$}}
\newcommand{\Db}{{\bar{D}}}

\newcommand{\hf}{\frac12}
\begin{document}                        
\begin{titlepage}

\begin{flushright} August, 2015 \\
Revised version: September, 2015
\end{flushright}

\vspace{5mm}

\begin{center}
{\large \bf  Complex linear Goldstino superfield and 
supergravity}
\end{center}
\begin{center}
{\large  
{Sergei M.\ Kuzenko}

\vspace{5mm}

\footnotesize{
{\it School of Physics M013, The University of Western Australia\\
35 Stirling Highway, Crawley W.A. 6009, Australia}}  

\vspace{2mm}}
\end{center}

\vspace{5mm}

\pdfbookmark[1]{Abstract}{abstract_bookmark}
\begin{abstract}
\baselineskip=14pt
The complex linear Goldstino superfield 
was proposed in arXiv:1102.3042 for the cases of global and local
four-dimensional $\cN=1$ supersymmetry. 
Here we make use of this superfield to construct a supergravity action
which is invariant under spontaneously broken local $\cN=1$ supersymmetry and has a positive cosmological constant for certain values of the parameters. 
\end{abstract}
\vfill

\end{titlepage}



\section{Introduction}

Four years ago, we constructed 
the Goldstino model  \cite{KT} described by 
a complex linear superfield $\S$ constrained by
\begin{align} \label{mCL:constraint}
	-\frac14 \Db^2 \S = f\,, \qquad f = {\rm const}\,.
\end{align}
Here $f$ is a parameter of mass dimension 2 which, 
without loss of generality, can be chosen to be real.
To describe the Goldstino dynamics, $\S$ was subject to 
the nonlinear constraints:
\begin{align}
	\S^2 &= 0~, \label{1st constraint} \\
	-\frac{1}{4} \S\Db^2D_\a\S &= f D_\a\S~. \label{2nd constraint}
\end{align}
The constraint \eqref{1st constraint} means that $\S$ is nilpotent.
The constraints \eqref{mCL:constraint}, \eqref{1st constraint} and 
\eqref{2nd constraint} imply that all component fields of $\S$ are constructed 
in terms of a single spinor field ${\bar \r}^\da$.
We recall that the general solution to the  constraint \eqref{mCL:constraint} is
\begin{align} \label{mCL:components}
	\S (\q,\bar\q)= \rme^{\rmi\q\s^a\bar\q\pd_a}
		\left(\f + \q\j + \sqrt2\bar\q\bar\r
		+ \q^2F + \bar\q^2f + \q^\a\bar\q^\da U_\ada + \q^2\bar\q\bar\c
		\right)\ .
\end{align}
The general solution to the constraint  \eqref{1st constraint}
fixes $\f$ and two of the auxiliary fields
\begin{align} \label{soln-to-1st-constraint}
	f \f   = \frac12\bar\r^2\,, \quad
	f\j_\a = \frac1{\sqrt2} U_\ada\bar\r^\da\,, \quad
	f F    = \frac1{\sqrt2}\bar\c\bar\r + \frac14 U^a U_a \ .
\end{align}
Finally, taking into account the constraint \eqref{2nd constraint} 
fixes all of the components as functions of the Goldstino $\bar\r$. 
The explicit expressions are:
\begin{equation}\label{solns-to-both-constraints}
\begin{gathered}	f \f			= \frac12\bar\r^2\,, \quad
	\sqrt2f^2\j_\a	= -\rmi\bar\r^2(\pd\bar\r)_\a \,, \quad
	f^{3}F 			= \bar\r^2(\pd_a\bar\r\st^{ab}\pd_b\bar\r)\,, \\
	f U_\ada		= 2\rmi(\s^a\bar\r)_\a\pd_a\bar\r_\db\,,  \quad
	f^2\bar\c_\da 	= \sqrt2\big((\bar\r\st^a\s^b\pd_b\bar\r)\pd_a\bar\r_\db
						- \frac12 (\Box\bar\r^2)\bar\r_\db \big)\ .
\end{gathered}\end{equation}

The form of the Goldstino action coincides with the free action for the complex linear superfield,
\bea
	S[\S,\bar\S] 
	= - \intx\!\rmd^2\q\rmd^2\bar\q\, \S\bar\S  ~.
	\label{Gaction}
\eea
At the component level,  
this action was shown \cite{KT}  to be the same as 
the one described by Samuel and Wess \cite{SamuelWess1983}.
 A nonlinear field redefinition relating the component form of \eqref{Gaction} to
  the Volkov-Akulov action \cite{VA} follows from the results in 
 \cite{KT0}.

It was also shown in \cite{KT} that all known Goldstino superfields 
\cite{SamuelWess1983,Rocek,IK2,LR}
can be obtained as composites constructed from spinor covariant derivatives of
$\S$ and its conjugate (see also \cite{FHKvU,KT2}).\footnote{The Goldstino 
superfields constructed in \cite{KT,SamuelWess1983,Rocek,LR} can be derived using 
the general relationship between linear and nonlinear realisations of supersymmetry
established in \cite{IK1,IK2}. In particular, the spinor Goldsino superfield 
advocated in \cite{SamuelWess1983} was first constructed in \cite{IK2}.}
This property and the universality  \cite{VA,Ivanov,Uematsu:1981rj}
of the Goldstino \cite{VA} implies that any model for global supersymmetry breaking 
can be described in terms of $\S$ and its conjugate.

Couplings of the complex linear Goldstino superfield to 
supersymmetric matter and $\cN=1$ supergravity were given in \cite{KT}. 
The results of \cite{KT} make it possible to derive 
 a simple construction of models for spontaneously broken 
 $\cN=1$ supergravity, similar to the old chiral construction of 
 \cite{LR}. Recently, there have appeared 
models for  spontaneously broken local 
supersymmetry \cite{BFKVP,HY}, which are 
based on the use of the  chiral Goldstino superfield proposed in \cite{Casalbuoni1989,KS}.
Here we demonstrate how the complex linear Goldstino of \cite{KT} 
can be used to describe spontaneously broken supergravity.


\section{Coupling to supergravity}

The supergravity generalisation\footnote{Our conventions for $\cN=1$ supergravity mainly correspond to \cite{BK}, 
with the only exception that  the full superspace integration measure $E$ in 
\eqref{omsg} is denoted $E^{-1}$ in \cite{Ideas}.}
of the constraints \eqref{mCL:constraint} and \eqref{2nd constraint} given in \cite{KT}
is
as follows:
\begin{subequations} \label{16.0}
\begin{align}
\label{mCL:constraint SuGra}
 	-\frac14 (\cDb^2 - 4R) \S &= Y\,, \qquad 	\cDb_\da Y = 0\,,  \\
\label{2nd constraint SuGra}
	-\frac14 \S  (\cDb^2 - 4R) \cD_\a\S &= Y \cD_\a\S\ , 
\end{align}
\end{subequations}
for some covariantly chiral scalar $Y$.\footnote{In the super-Poincar\'e  case, 
modified linear constraints of the form 
\eqref{mCL:constraint SuGra} were first introduced in \cite{DG}.}  
Here $\cD_A =(\cD_a , \cD_\a , \cDb^\da)$ 
denote the superspace covariant derivative corresponding to 
the Wess-Zumino formulation for $\cN=1$ supergravity \cite{WZ} 
(which at the component level is equivalent to approaches 
developed in \cite{old}), 
and $R$ is the covariantly chiral scalar component of the superspace torsion  
described in terms of $R$, $G_{\a \da}$ and $W_{\a \b \g}$
(see \cite{Ideas} for a review). 
Of course, the constraints \eqref{mCL:constraint SuGra} and 
\eqref{2nd constraint SuGra} 
have to be accompanied by the nilpotent condition \eqref{1st constraint}. 

Here we consider the simplest case when $Y$ is a real non-zero constant, 
\bea
Y = f  =  \const~, \label{17}
\eea 
as in the rigid supersymmetric case.
To describe the dynamics of the Goldstino superfield coupled 
to supergravity, we propose the following action:
\bea
\label{omsg}
S = - 
\int {\rm d}^
{4} x \rmd^2\q\rmd^2\bar\q\,
E\left( \frac{3}{\bm \k^2} + \bar \S  \S \right)
+ \left\{ \frac{\bm \m}{\bm \k^2} \int {\rm d}^
{4} x \rmd^2\q\,
\cal E  + {\rm c.c.} \right\}
~, 
\eea
where $\bm \k$ is the gravitational coupling constant and  
$\bm \m$ a cosmological parameter.  The integration measures
 $E$ and $\cal E$ correspond to  the full superspace 
and its chiral subspace, respectively.  
The action \eqref{omsg} describes old minimal $\cN=1$ supergravity 
if  $\S$ and $\bar \S$ are switched off. 
With the Goldstino superfields $\S$ and $\bar \S$
included, the action proves to describe spontaneously broken $\cN=1$ supergravity.

\section{Super-Weyl invariant reformulation} 

To reduce the action \eqref{omsg} to components, one can follow, e.g., the 
component reduction procedure described in \cite{Ideas}. 
A less tedious calculation is required if one introduces a super-Weyl invariant extension 
of  \eqref{omsg} by coupling this theory to a covariantly chiral conformal compensator 
 $\F$, $\bar{\cal D}_\da \F=0$, which is assumed to be nowhere vanishing. 
Such an extension can be obtained 
following the scheme described in \cite{KMcC} and based 
on the ideas due to Kugo and Uehara \cite{KU}. 

The super-Weyl transformation \cite{HT} in $\cN=1$ old minimal supergravity is
\bea
{\cal D}'_\a = {\rm e}^{ \hf \s - {\bar \s} } 
\Big( {\cal D}_\a - ({\cal D}^\b \s) \, M_{\a \b} \Big) ~, \qquad
\bar{\cal D}'_\da ={\rm e}^{ \hf {\bar \s} - \s } \Big(
\bar{\cal D}_\da -  (\bar{\cal D}^\db {\bar \s}) {\bar M}_{\db\da} \Big)~,
\label{superweyl}
\eea
where $\s $ is an arbitrary  covariantly chiral scalar parameter,
$\bar{\cal D}_\da \s=0$. 
The super-Weyl transformation of the chiral compensator $\F$ is 
\bea
\F' = {\rm e}^{-\s} \F~.
\eea
A super-Weyl invariant extension of the constraints \eqref{16.0} 
with  $X$ given by \eqref{17} is
\begin{subequations} \label{23}
\begin{align}
 	-\frac14 (\cDb^2 - 4R) \S &= f \F^2\,,  \label{23a}  \\
	-\frac14 \S  (\cDb^2 - 4R) \cD_\a( {\S}\bar \F^{-1})  &= f \F^2 \cD_\a
	({\S}\bar \F^{-1})\ , 
\end{align}
\end{subequations}
provided the super-Weyl transformation of $\S$ is chosen to be
\bea
\S' =  {\rm e}^{-\bar \s} \S~.
\eea
In the super-Weyl gauge $\F=1$, the constraints \eqref{23} reduce to 
\eqref{16.0}.\footnote{Applying a field redefinition 
$\S \to =\F^n \S$ leads to a different super-Weyl transformation law
and modifies the explicit form of constraints  \eqref{23}.}  

The super-Weyl invariant extension of the action \eqref{omsg} is
\bea
S = - 
\int {\rm d}^
{4} x \rmd^2\q\rmd^2\bar\q\,
E\left( \frac{3}{\bm \k^2} \bar \F \F + \bar \S  \S \right)
+ \left\{    \frac{\bm \m}{\bm \k^2} 
\int {\rm d}^{4} x \rmd^2 \q\,
{\cal E} \,\Phi^3 
+ {\rm c.c.} \right\}
~.
\label{155}
\eea
This action can be  reduced to components using the reduction formula 
(5.8.50) in \cite{Ideas} 
and imposing suitable super-Weyl gauge conditions on the components
of the chiral compensator $\F$, following the patterns described in \cite{KMcC} 
for four-dimensional  $\cN=1$ supergravity-matter systems 
and also in \cite{KLRST-M} for three-dimensional $\cN=2$ supergravity-matter theories.  
Upon reducing the action to components 
and eliminating the supergravity auxiliary fields, 
for the cosmological constant one obtains
\bea
\Lambda = f^2 - 3\frac{|{\bm \mu} |^2}{{\bm \k}^2}~.
\label{16}
\eea
This value agrees with the recent results in \cite{ADFS,BFKVP,HY},\footnote{Actually,  
the explicit expression 
\eqref{16} for the cosmological constant can be obtained without any calculation.
The second term in \eqref{16} is the standard cosmological constant in pure $\cN= 1$ supergravity \cite{Townsend}, see, e.g., section 6.1.4 in \cite{Ideas} for a review.
The first term in \eqref{16} follows from eq. (2.4) in \cite{KT}.} 
as well as with the ancient results \cite{super-Higgs,LR} (see also \cite{CGP}).
The cosmological constant is positive for ${{\bm \k}^2} f^2 > 3{|{\bm \mu} |^2}$.
Since the theory possesses local supersymmetry, the Goldstino can be eaten 
by the gravitino, in accordance with the Higgs mechanism for local supersymmetry 
\cite{VS,super-Higgs} (known as the super-Higgs effect). 
As a result, the gravitino becomes massive.

In conclusion, we recall that the constraint \eqref{mCL:constraint}
is the only way to describe $\cN=1$ anti-de Sitter supergravity
using a non-minimal scalar multiplet as  compensator \cite{BK}.
We have shown that the same constraint can be used 
to describe  spontaneously broken $\cN=1$ supergravity 
with a positive cosmological constant. 

\section{Conclusion}

Unlike the Goldstino superfields introduced in \cite{SamuelWess1983,Rocek,IK2,LR},
the complex linear Goldstino superfield allows nontrivial matter couplings \cite{KT}.
This is achieved by choosing the chiral scalar $Y$ in \eqref{16.0} to be 
\bea
Y = Y(\varphi^i)~, 
\eea
for some matter chiral superfields $\varphi^i$. 
In the presence of matter, the action \eqref{omsg} is replaced with 
\bea
S = &-& 
\int {\rm d}^
{4} x \rmd^2\q\rmd^2\bar\q\,
E\left( \frac{3}{\bm \k^2} {\rm e}^{-\frac{1}{3} K\!(\varphi,{\bar \varphi})}
+ \bar \S  \S \right) \non \\
&&\qquad \qquad +  \left\{  
\int {\rm d}^{4} x \rmd^2 \q\,
{\cal E} \,\Phi^3 \Big(  \frac{\bm \m}{\bm \k^2}   + W(\varphi)  \Big)
+ {\rm c.c.} \right\}~,
\eea
with $K\!(\varphi,{\bar \varphi})$ being the K\"ahler potential of a K\"ahler manifold
and $W(\varphi) $ a superpotential. Such matter couplings are analogous 
to the models considered in \cite{BFKVP,HY,ADFS}.

If $\S$ obeys {\it only} the constraint \eqref{23a}, 
the supergravity-matter system \eqref{155} possesses 
a dual formulation \cite{KT,BK}
\bea
S = - 
\int {\rm d}^
{4} x \rmd^2\q\rmd^2\bar\q\,
E\left( \frac{3}{\bm \k^2} \bar \F \F- \bar X  X \right)
+ \left\{  
\int {\rm d}^{4} x \rmd^2 \q\,
{\cal E} \,\Phi^3 \Big(  \frac{\bm \m}{\bm \k^2}   +f \frac{X}{\F}  \Big)
+ {\rm c.c.} \right\}
~, ~~~
\label{SUGRA17}
\eea
where  $X$ is a  chiral scalar superfield, $\bar{\cal D}_\da X=0$,
 with the super-Weyl transformation 
\bea
X' = {\rm e}^{-\s} X~.
\eea
It is not clear to us how to modify the duality transformation 
in order to take account of the nilpotent constraint 
\eqref{1st constraint}. 

The action \eqref{SUGRA17} with the chiral scalar $X$ 
constrained by 
\bea
X^2 =0~
\label{X-constraint}
\eea
describes coupling to supergravity of the Goldstino superfield introduced in 
\cite{Casalbuoni1989,KS}. This model for spontaneously broken supergravity has been of much 
interest recently \cite{BFKVP,HY,ADFS,DFKS,AM}, in particular since it admits a 
nice geometric reformulation  \cite{ADFS,DFKS,AM}. Here we would like to present a slightly different derivation of such a reformulation. In what follows, $X$ is assumed to obey the nilpotent constraint \eqref{X-constraint}.

Varying \eqref{SUGRA17} with respect to $\F$ gives the equation
\bea
{\mathbb R} -{\bm \m} = \frac{2}{3} f \k^2 \frac{X}{\F} ~,
\label{Phi-equation}
\eea
where we have the super-Weyl invariant chiral scalar 
\bea
{\mathbb R} = -\frac{1}{4} \F^{-2} (\cDb^2 - 4R) \bar \F~,
\eea
which is related to the chiral scalar $\cal R$ of \cite{DFKS} by the rule 
${\cal R} = \F \mathbb R$.
Due to the nilpotent constraint \eqref{X-constraint}, the equation of motion 
\eqref{Phi-equation} implies that 
\begin{subequations}\label{2444}
\bea
({\mathbb R} -{\bm \m})^2 =0~, 
\label{244}
\eea
which has the same form as the constraint put forward in  \cite{DFKS,AM}.
Making use of \eqref{Phi-equation}  once again, 
the action  \eqref{SUGRA17} takes the geometric form 
\bea
S = \Big( \frac{3}{2f \k^2}\Big)^2  
\int {\rm d}^{4} x \rmd^2\q\rmd^2\bar\q\,
E \, \bar \F \F|{\mathbb R} -{\bm \m} |^2
- \left\{  \hf  \frac{\bm \m}{\bm \k^2} 
\int {\rm d}^{4} x \rmd^2 \q\,
{\cal E} \,\Phi^3 
+ {\rm c.c.} \right\}
~,
\label{SUGRA23}
\eea
\end{subequations}
where $\mathbb R$ is subject to the constraint \eqref{244}.
This action differs in its functional form from the one proposed in  \cite{DFKS,AM}.
The latter coincides with the minimal supergravity action 
\begin{subequations} \label{2555}
\bea
S = -  \frac{3}{\bm \k^2}
\int {\rm d}^
{4} x \rmd^2\q\rmd^2\bar\q\,
E \,\bar \F \F
+ \left\{   \frac{{\bm \m}}{\bm \k^2} 
\int {\rm d}^{4} x \rmd^2 \q\,
{\cal E} \,\Phi^3    
+ {\rm c.c.} \right\}
~, ~~~
\label{SUGRA25}
\eea
where $\mathbb R$ is subject to the nilpotent  constraint 
\bea
({\mathbb R} -{\bm \l})^2 =0~, \qquad {\bm \l} =   \frac{{\bm \m}}{\bm \k} 
+\frac{1}{\sqrt{3} } {\bm \k} f~.
\eea 
\end{subequations}
The two descriptions should be equivalent. 

An interesting problem is to understand whether our model \eqref{155} admits a geometric formulation similar to  \eqref{2444} or \eqref{2555}.
\\

\noindent
{\bf Acknowledgements:} The author is grateful to Simon Tyler for discussions
and comments on the manuscript. Discussions with Daniel Butter and Evgeny Ivanov 
are also gratefully acknowledged. 
This work was supported by the ARC projects DP140103925.


\begin{footnotesize}
\providecommand{\href}[2]{#2}
\begingroup\raggedright
\endgroup
\end{footnotesize}

\end{document}